\documentclass[letterpaper,11pt]{article}
\usepackage{url,amsmath,amssymb,verbatim,setspace}
\usepackage{graphicx}
\usepackage[round,authoryear]{natbib}
\begin{document}

\title{Feature selection in simple neurons: how coding depends on spiking dynamics}

\author{Michael Famulare\\
        University of Washington\\
        Department of Physics \\
        Box 351560 \\
        Seattle, WA 98195-1560  \\
        \texttt{famulare@u.washington.edu}\\
        \\
        Adrienne Fairhall\\
        University of Washington\\
        Department of Physiology and Biophysics \\
        HSB G424, Box 357290\\
        Seattle, WA 98195-7290}

%\email{famulare@u.washington.edu}
%\affiliation{Department of Physics, University of Washington}

%\author{ Adrienne Fairhall\\
%        University of Washington\\
%        Department of Physiology and Biophysics \\
%        HSB G424, Box 357290\\
%        Seattle, WA 98195-7290}

%\affiliation{Department of Physiology and Biophysics, University of Washington}

\date{\today}

\maketitle

%\begin{center}
%\underline{Abbreviated title}:  Feature selection in simple neurons
%\end{center}

%\doublespacing

%\newpage

\begin{abstract}
{The relationship between a neuron's complex inputs and its spiking output defines the neuron's coding strategy. This is frequently and effectively modeled phenomenologically by one or more linear filters that extract the components of the stimulus that are relevant for triggering spikes, and a nonlinear function that relates stimulus to firing probability.  In many sensory systems, these two components of the coding strategy are found to adapt to changes in the statistics of the inputs, in such a way as to improve information transmission. Here, we show for two simple neuron models how feature selectivity as captured by the spike-triggered average depends both on the parameters of the model and on the statistical characteristics of the input.}
\end{abstract}

%\maketitle

%\section{Introduction}
Neuronal dynamics are characterized by nonlinearities that lead to large, approximately stereotyped voltage excursions, or spikes, that are the basis for interneuronal signaling. Capturing the relationship between inputs and the resulting pattern of spike outputs from a given neuron in the form of a reduced functional model is a focus of sensory neuroscience. In the sense that such a model provides a general mapping from input to output, it can be thought of as the neuron's ``coding strategy''.

Reverse correlation methods \citep*{Bryant1976,deBoer1968,Sakai1992,Hunter1986} provide a means to sample the statistical characteristics of stimuli that tend to trigger spikes; in the simplest case, the mean, or spike-triggered average stimulus (STA), is the optimal linear kernel for predicting the firing rate from the stimulus \citep*{RiekeSpikes}.
Using reverse correlation, one may obtain an approximate functional model for the neuronal input/output transformation in terms of the input features that drive the system \citep*{Meister1999,deRuyter1988,Simoncelli2004}. These methods may be applied not only to determine how neural systems are driven by external stimuli, but to extract a model for how specific patterns of synaptic current inputs drive single neurons. This allows one to determine the role that a single neuron with a characteristic complement of ion channels plays in a circuit: the integration of inputs over a certain timescale \citep*{Slee2005,Svirskis2003,Prescott2006}, the detection of sudden change or highly synchronous events \citep*{Abeles1982,Slee2005,Svirskis2004}, or the selection of certain frequency components in the input \citep*{Izhikevich2001,Prescott2006}.

Here, we will derive explicit expressions for the outcome of such a statistical analysis applied to two simple neuron models. We have two goals. The first is to develop a general framework for understanding how the details of neuronal dynamics establish or influence the features in the input that trigger spikes. Second, neuronal systems show adaptation to statistics, in the sense that the neuron's coding strategy often changes when driven by stimuli with different statistical properties. In the case of single neurons, such effects can modulate or gate the effective computation of the neuron according to the statistical properties of the signal or the background inputs \citep*{Hasenstaub2007,Destexhe1999,Fellous2003}. To identify the rules governing this process, one would like to know to what extent the observed changes may result from time-independent neuronal nonlinearities and to what extent they must be due to changes in underlying neuronal parameters. To study this, we will compute how the experimentally obtained features of two fixed models depend on the statistical properties of the stimulus, focusing on the variance of a white noise input.

The key points of this paper are:
\begin{itemize}
\item The relevant linear filter corresponding to a nonlinear spiking neuron model is determined by the nonlinear dynamics linearized in a manner consistent with the typical operating regime of the system, which is determined both by its dynamics and by the stimulus conditions.
      To characterize this regime, we compute the voltage probability distributions from the Fokker-Planck interpretation of the models.
\item We then use a novel application of the technique of \textit{stochastic linearization} to map the nonlinear models onto a set of linear models.  By studying both the mapping, determined by an optimization function relating the linear and nonlinear models, and the related STA predictions for the equivalent linear system, we can delineate the roles of different nonlinearities on spike encoding.
\item The form of the STA is influenced both by the subthreshold (non)linear dynamics and the spike afterhyperpolarization.
\item Models with similar phase space topology can have STAs whose form is controlled by different mechanisms.  A rapid-onset exponential integrate-and-fire model (EIF) has no significant subthreshold nonlinearity, and so its STA is almost completely determined by the probability current due to spiking.  In contrast, the quadratic integrate-and-fire model (QIF), while superficially similar to the EIF, has an STA whose form is dominated by the sampling of the subthreshold nonlinearity, with spiking effects playing a secondary role.
\end{itemize}

\section{Models and numerical methods}
\label{sec:model}
Change in the effective feature selectivity with driving variance has been studied for the case of the leaky integrate-and-fire (LIF) model \citep*{Paninski2003FP,Paninski2006IFSTA,Yu2003}. In the LIF model, the dynamics are linear until the voltage reaches an imposed threshold after which the voltage is immediately reset below threshold. Thus, the LIF contains no intrinsic excitability, and further, does not allow for the possibility that the system can cross threshold multiple times before spiking due to noisy inputs. This discontinous behavior with respect to spike initiation is not found in biological neurons.

Two simple models with more realistic spike initiation are the quadratic and exponential integrate-and-fire models (QIF and EIF, respectively) \citep*{Ermentrout1986,Fourcaud2003}. Both models are similar in spirit to the LIF insofar as they replace the afterhyperpolarization mechanism with a discontinous jump, or after-spike reset, but the point of reset in these models occurs at the peak of the spike instead of at the threshold voltage.  This mitigates the effects of the pathological behavior in response to noise that the discontinuity creates \citep*{Paninski2006IFSTA} by moving it away from the interesting region of spike initiation.  The models are described by an equation of the form:
\begin{equation}
\tau_m\dot{v}=-v+ f(v) +\left(V_{r}-V_{s}\right)\delta(v-V_{s}) + s(t),
\label{eq:model}
\end{equation}
where $v$ denotes the membrane voltage, $\tau_m$ is the membrane time constant, $V_{s}$ is the voltage that defines the spike height and $V_{r}$ is the post-spike reset voltage. The input current, $s(t)$, is a zero-mean gaussian white noise (GWN) process with correlation function $\langle s(t)s(0) \rangle=\sigma^2\tau_m\delta(t)$.    The delta-function is shorthand for the act of resetting the voltage to $V_r$ after it reaches $V_s$.  All of the spike-generating and nonlinear subthreshold dynamics are encoded in $f(v)$.  For the two models studied here, we have:
\begin{equation}
f(v) = \begin{cases} \alpha v^2 & \text{for the QIF,} \\ g\exp\left[\frac{v-V_*}{g}\right] & \text{for the EIF.} \end{cases}
\end{equation}
We study parameters such that the resting potential is zero and the unstable fixed point is at $\alpha^{-1}$ for both models.  This requires us to choose $V_*=\alpha^{-1}\left(1+g\alpha\ln(g\alpha)\right)$ and $g\ll V_*$.  Somewhat paradoxically, despite the higher order nonlinearity, the choice of small $g$ causes the exponential nonlinearity to turn on over a much tighter range in voltage than the quadratic nonlinearity of the QIF.  We will see that this leads to more linear behavior of the EIF model below threshold. Thus the two models behave noticeably differently below threshold while still having the same after-spike dynamics.

\subsection{Reverse correlation analysis}
Reverse correlation is used to determine characteristics of the stimulus that are correlated with neuronal response. From a long, random stimulus presentation $s(t)$ and the resulting spike response times $t_i$, one collects the set of $N$ current traces that led to a spike, $s(\tau-t_i)$, over an interval of time $\tau = [0,-T]$ prior to the spike where $T$ is chosen appropriately to capture all the stimulus history that is relevant to triggering the spike.  The {\em spike-triggered average} or STA, $\bar{s}(\tau)$, is found by averaging these samples over $i$:
\begin{equation}
\bar{s}(\tau) = \frac{1}{N} \sum_{i=1}^N s(\tau - t_i).
\label{eq:sta}
\end{equation}
\subsection{Defining spike times}
The results of reverse correlation analysis can depend on how the spike time is defined.  Here, we will look at the STAs with a temporal resolution that is short compared to the average spike width.  Different choices of the voltage threshold used to define spikes will accordingly lead to STAs that differ from each other due to temporal jittering of the ensemble of spike-triggered trajectories.
Because we want to understand how the spiking of the model determines the feature selected from the stimulus ensemble, we are interested in choosing a threshold that yields an STA that best captures the role of the stimulus on the approach to the spike but is not sensitive to stimulus-driven variations in the spike itself.  For both models considered here, this is achieved by selecting the unstable fixed point that separates the subthreshold region from the spiking region in the absence of noise.
Since the location of the unstable fixed point is a function of the mean input current and the quadratic form of the nonlinearity, we will call it the {\em dynamical threshold} in accordance with previous work \citep*{Izhikevich2000,Hong2007}.  For the zero-mean inputs considered here, the dynamical threshold is $V_{th}=\alpha^{-1}$.
Thus, we define spike times as the time of the last upward crossing of the dynamical threshold preceding an after-spike reset.

\subsection{Model simulation}
In discrete-time with time step $h$, the nonlinear models in equation \ref{eq:model} were realized as:
\begin{eqnarray}
v_{n+1}=v_n+\frac{h}{\tau_m}\left(-v_n+f(v_n)\right)+\sigma\sqrt{\frac{h}{\tau_m}}\xi_n\ , \label{eq:QIFdiscreet}\\
\text{if } v_{n+1}\ge V_{s}, \text{ then } v_{n+1}\rightarrow V_{reset}\ .
\end{eqnarray}
where the $\xi_n$ are drawn from a gaussian distribution with zero mean and unit variance.
For all figures in this paper, simulations were run with a time step of $h=\tau_m/200$ until $2 \times 10^5$ spikes were accumulated.  The noise was generated with \texttt{randn} in Matlab R2007b.  Parameters used in simulation: $\alpha=1$, $\tau_m=1$, $V_{s}=25$, $V_{r}=-0.2$, $g=\frac{1}{10}$, and $V_*=0.77$.

\section{Numerical results}
\label{sec:numerics}
\begin{figure}[!ht]
\centering
\includegraphics[width=4.25in]{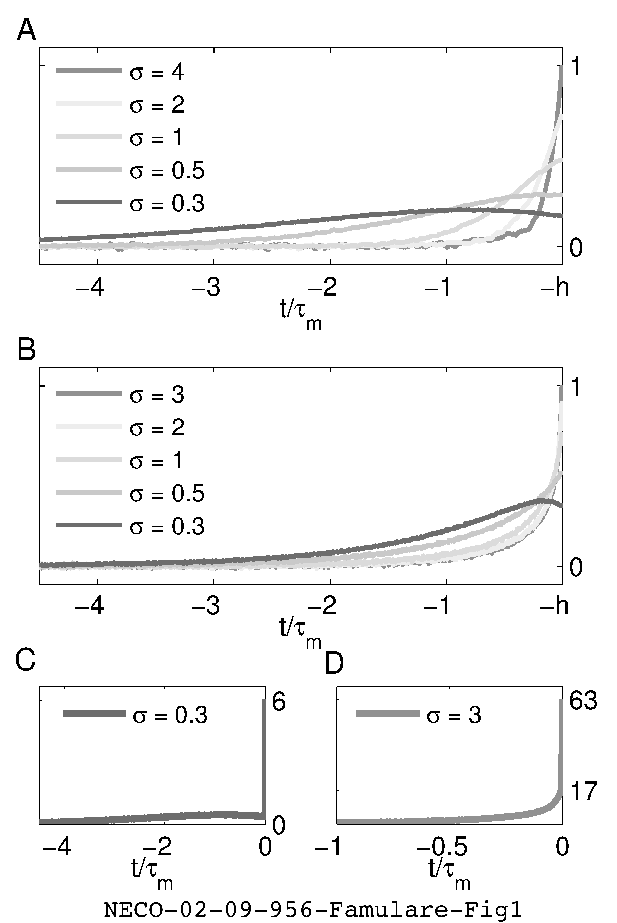}
\caption{The STAs for (A) the QIF and (B) the EIF (using the L2-norm for easy visual comparison), triggered on $V_{th}=\alpha^{-1}$, for various $\sigma$ with the upward step at $t=0$ removed.  Note that at small $\sigma$, the STA is non-monotonic while, at large $\sigma$, it is approximately a decaying exponential.  As representative examples,  the STA in real units is shown (C) for the QIF for $\sigma=0.3$ and (D) for the EIF for $\sigma=3$ with the last time-step included.}
\label{fig:STAnum}
\end{figure}

We computed the STA numerically for a range of values of the stimulus standard deviation $\sigma$ for both models.
Results are shown in figure \ref{fig:STAnum}. The STA at all values of $\sigma$ has two components: an extended feature and a sharp upward step at the time of the spike.  For the feature, two different types of behavior appear. For large $\sigma$, the STAs are approximately decaying exponentials for which, as the standard deviation increases, the decay timescale decreases and the amplitude increases. Thus, at larger $\sigma$, the QIF and EIF models perform approximately linear leaky integration, where the effective leakiness depends on the standard deviation.  For very small $\sigma$, the STAs are non-monotonic, with the peak amplitude occurring well before the spike time.

\section{Approximate STA for finite standard deviations}
\label{sec:stochlin}
To best understand how details of the models influence the STAs, we would like to be able to calculate the STAs analytically. In the zero standard deviation limit, the STA can be analytically calculated for the QIF via a large deviations principle and path integral methods \citep*{Paninski2006STV,Badel2008a,Wilson2008}, but that type of analysis does not extend to finite $\sigma$.   However, path integral methods can be applied for arbitrary $\sigma$ to perfectly linear models with no reset.  As noted previously \citep*{Hong2007}, the observation that the STA is an exponential implies that the subthreshold dynamics of the model are effectively linear.  Since the STAs of the nonlinear models are roughly exponential for larger standard deviations, we should be able to introduce a linear approximation to the nonlinear models that captures the qualitative behavior of the STA and helps explain in detail how the STA arises from the form of the nonlinearity.

The main idea is as follows.  While it is impossible to derive complete, time-dependent statistical distributions for these models, we can get the steady-state distribution from the Fokker-Planck equation. This distribution gives us information about how the properties of the stimulus and spiking dynamics determine how the system samples its subthreshold nonlinearities. We will then use the steady-state distribution to map the nonlinear models onto linear models and thus compute an approximation to the STA.  Since the linear model follows from the steady-state distribution, we can think of the linear model as describing the ``time-averaged dynamics'' of the nonlinear models.

\subsection{The steady-state distribution}
A key ingredient for understanding the behavior of the spiking models and for determining an analytically tractable mapping of a nonlinear model to a linear model is the steady-state probability distribution for the voltage in response to an input with given statistical characteristics.
This probability distribution, $p_N(v)$, can be computed from the Fokker-Planck equation \citep*{Paninski2003FP,Brunel2003,Lindner2003,Fourcaud2003}, which for models of the form given in equation \ref{eq:model} is:
%\begin{comment}
\begin{eqnarray}
\frac{\partial p_N(v,t)}{\partial t}&=&\frac{\partial}{\partial v}\left[\left(\frac{v-f(v)}{\tau_m}\right) p_N(v,t)\right]+\frac{\sigma^2}{2\tau_m}\frac{\partial^2p_N(v,t)}{\partial v^2} \nonumber \\
&& \qquad +R(t)\left[\delta(v-V_r)-\delta(v-V_{s})\right],
\end{eqnarray}
%\end{comment}
where $R(t)$ is the time-dependent mean firing rate that needs to be determined self-consistently in solving the equation.
This is a continuity equation for $p_N\left(v, t\right)$ which expresses that the evolution of the distribution is driven by the deterministic nonlinear driving force, diffusion, and spiking.
We are interested in the steady-state distribution, for which $\frac{\partial p_N}{\partial t}=0$ and $R(t)$ goes to the mean rate $R$.  Using standard methods \citep*{Risken}, one can show that the steady state distribution is:
\begin{equation}
p_N(v)=\frac{2R\tau_m}{\sigma^2}\displaystyle e^{\frac{-1}{\sigma^2}\left(v^2-2F(v)\right)}\int_{\max(v,V_r)}^{V_s}dv'\, e^{\frac{1}{\sigma^2}\left(v'^2-2F(v')\right) },
\label{eq:P(v)}
\end{equation}
where $F(v)=\int f(v)dv$, and the mean firing rate is the normalization constant.

This distribution is the product of a Boltzmann factor, controlled entirely by the nonlinear dynamics, $F(v)$, preceding a spike, and a spiking flux term which carries the dependence on the spike parameters $V_r$ and $V_s$. Since we are mainly interested in behavior below the unstable fixed point, or dynamical threshold, and the models considered here have reset voltages, $V_r$, near the resting potential, the contribution of the spiking flux term does not depend strongly on the form of $F(v)$, but does depend strongly on the location of $V_r$.

\subsection{Stochastic linearization}
We turn to a set of techniques known as {\em stochastic linearization} (SL) (see \citep*{Socha2005p1} for an extensive review) to model and understand the behavior of the STA of the nonlinear models. In the SL approach, one seeks the parameters of a linear model that optimally capture the properties of the nonlinear model in a regime of interest.  In our case, we are interested in the linear model that best captures the approach of a nonlinear model to threshold for a given input standard deviation, but we are unconcerned with the dynamics of the spike itself. Thus, we search for an equivalent linear model of the form:
\begin{equation}
\tau_m \dot{v}=-k_{\sigma}v+c_{\sigma}+s(t),
\label{eq:linear}
\end{equation}
where we make no attempt to model the spike or the reset \citep*{GerstnerKistler}.  This linear model is simply an Ornstein-Uhlenbeck process \citep*{Risken}, and it has the associated steady-state probability distribution:
\begin{equation}
p_L(v)=\sqrt{\frac{k_{\sigma}}{\pi\sigma^2}}\text{exp}\left[-\frac{k_{\sigma}}{\sigma^2}\left(v-\frac{c_{\sigma}}{k_{\sigma}}\right)^2\right]. \label{eq:LinStead}
\end{equation}

To determine the parameters of the optimal linear model, we must select an optimization function that maps the nonlinear model onto the linear model.  There are no unique methods for choosing optimization functions that will yield good results \citep*{Socha2005p1}, and different functions will generally yield different results.  We focus on two optimization functions which give weight to different properties of the nonlinear model. 

\subsection{Minimizing the Kullback-Leibler divergence}
The Kullback-Leibler divergence ($D_{KL}$) measures the similarity of two probability distributions \citep*{CoverThomas}. To map the nonlinear models to sets of linear models, we can use the $D_{KL}$ to minimize the difference between the subthreshold part of the nonlinear steady-state distribution and the matched linear model's steady-state distribution.  The $D_{KL}$ for this problem is:
\begin{eqnarray}
D_{KL}\left(p_N||p_L\right)&=&\displaystyle \int_{-\infty}^{V_{th}} dv \frac{p_N(v)}{Z_N}\ln\frac{p_N
(v)}{Z_Np_L(v)} ,\label{eq:DKL}\\
\text{where } Z_N&=&\int_{-\infty}^{V_{th}}p_N(v). \nonumber
\end{eqnarray}
To find the optimal linear model with this criterion, we minimize the $D_{KL}$ with respect to $k_{\sigma}$ and $c_{\sigma}$. Doing so yields
\begin{eqnarray}
k_{\sigma}&=&\frac{\sigma^2}{2\left(E\left[v^2\right]-E\left[v\right]^2\right)}, \nonumber \\
c_{\sigma}&=&k_{\sigma}E\left[v\right], \label{eq:DKLkc}\\
\text{where } E\left[\ldots\right]&=&Z_N^{-1}\displaystyle \int_{-\infty}^{V_{th}}p_N(v)\left[\ldots\right]. \nonumber
\end{eqnarray}
In the $\sigma\rightarrow 0$ limit, $k_0=1$ and $c_0=0$, corresponding to the classical linearization around the fixed point of a nonlinear model.  These expressions show that minimizing the $D_{KL}$ amounts to simply estimating the mean and variance below threshold.  This criterion is only sensitive to the probability distribution itself and has no knowledge of the underlying dynamics.  

\subsection{Minimizing the energy below threshold}
An alternative optimization criterion is to optimize the mean square error in the energy, or first integral of the nonlinear models, below threshold.  The energy of the nonlinear model is
$$ E=\frac{v^2}{2}-F(v), $$ and so the optimization criterion for $k_{\sigma}$ and $c_{\sigma}$ is
\begin{equation}
I=E\left[\left(\frac{v^2}{2}(1-k_{\sigma})+cv-F(v)\right)^2\right],
\label{eq:energy}
\end{equation}
where $F(v)$ and $E\left[\ldots\right]$ are defined as before.  This criterion amounts to trying to match the Boltzmann part of the distributions, taking spiking into account only through the bias it provides to the expectation value.  This piece is primarily sensitive to the specifics of the dynamics below threshold and is less sensitive to the overall shape of the distribution than the $D_{KL}$ is.  Minimizing $I$ yields:
\begin{eqnarray}
k_{\sigma}&=&1-2\frac{E\left[v^2F(v)\right]E\left[v^2\right]-E\left[vF(v)\right]E\left[v^3\right]}{E\left[v^4\right]E\left[v^2\right]-E\left[v^3\right]^2} \nonumber \\
c_{\sigma}&=&\frac{E\left[vF(v)\right]E\left[v^4\right]-E\left[v^2F(v)\right]E\left[v^3\right]}{E\left[v^4\right]E\left[v^2\right]-E\left[v^3\right]^2}\label{eq:Ikc}
\end{eqnarray}
Again, in the $\sigma\rightarrow 0$ limit, $k_0=1$ and $c_0=0$.  In this case, we see that the optimal parameters are directly sensitive to the form of the nonlinearity below threshold and that the shape of the probability distribution only enters through the expectation values.

\subsection{The meanings of the optimization criteria}

The two optimization criteria give different weights to different roles of the nonlinearity.  The $D_{KL}$ criterion is sensitive to the net statistical distribution below threshold, regardless of whether it comes about due to the spike or the subthreshold nonlinearity, whereas the energy criterion is primarily sensitive to the form of the subthreshold nonlinearity.  Accordingly, we can expect that the linear model, for a given nonlinear model and input standard deviation, found by the different criteria will be different.  Specifically, for nonlinear models whose optimal linear equivalents are best described by the energy criterion, the parameters $k_{\sigma}$ and $c_{\sigma}$ will be closely related to the form of the subthreshold nonlinearity but may not be very sensitive to the overall details of the voltage distribution below threshold.  In contrast, for models with linear equivalents that are best described by the $D_{KL}$ criterion, the parameters may have essentially no relation with the subthreshold nonlinearity, but rather describe global statistical properties set by the mean and variance of the voltage distribution.

\subsection{STA of the linear model}
To find the STA for the linear model and compare it to numerical simulations, we move to discrete time by defining $t=nh$, where $n$ is an integer and $h$ is the time step. For clarity of notation, we identify $v(t)=v(nh) \equiv v_n$. The linear model in equation \ref{eq:linear} is equivalently described by the forward transition probability distribution:
\begin{equation}
p\left(v_{n+1}|v_{n}\right)=\sqrt{\frac{\tau_m}{2\pi\sigma^2h}}\text{exp}\left[-\frac{\tau_m}{2\sigma^2h}\left(v_{n+1}-\left(1-\frac{hk_{\sigma}}{\tau_m}\right)v_{n}-\frac{hc_{\sigma}}{\tau_m}\right)^2\right].
\label{eq:Pvnvn1}
\end{equation}
Also of use are the steady-state probability distribution, $p_L(v)$, given in equation \ref{eq:LinStead}, and the backward transition probability distribution, $p(v_{n}|v_{n+1})$, which can be derived with Bayes' rule:
\begin{align}
p\left(v_n|v_{n+1}\right)&=\frac{p(v_{n+1}|v_{n})p_L\!\left(v_n\right)}{p_L\!\left(v_{n+1}\right)},\notag \\
&=\sqrt{\frac{\tau_m}{2\pi\sigma^2h}}\text{exp}\left[-\frac{\tau_m}{2\sigma^2h}\left(v_n-\left(1-\frac{hk_{\sigma}}{\tau_m}\right)v_{n+1}-\frac{hc_{\sigma}}{\tau_m}\right)^2\right], \label{eq:Pvn1vn}
\end{align}
for small $\frac{h}{\tau_m}$. Notice that the linear model is statistically reversible \citep*{Weiss1975}: the backward transition distribution is the time reversal of the forward, \mbox{$v_n\rightleftarrows v_{n+1}$}.
Since the model is linear, the STA follows from the spike-triggered voltage, $\bar{v}$, via:
\begin{eqnarray}
\bar{s}(t)&=&\tau_m\dot{\bar{v}}(t)+k_{\sigma}\bar{v}(t)-c_{\sigma}, \label{eq:STAdeflin}\\
\bar{s}_n&=&\left(\bar{v}_n-\bar{v}_{n-1}\right)\frac{\tau_m}{h}+k_{\sigma}\bar{v}_{n-1}-c_{\sigma}.
\end{eqnarray}
The spike-triggered voltages of the linear model can be found exactly with the following recipe.  We start at the spike time, $t=0 \ (n=0)$---the first time for which $v>V_{th}$ and $\dot{v}>0$.

The mean voltage at the spike time, $\bar{v}_0$, is given by:
\begin{eqnarray}
\bar{v}_0=\displaystyle \int_{-\infty}^{\infty} dv_0\,v_0\,p(v_0|{\rm spike})
\label{eq:STV0}
\end{eqnarray}
The spike-triggered voltage distribution, $p(v_0|{\rm spike})$, follows from the threshold-crossing condition.  The probability of finding a voltage $v_0$ at the spike time is given by the probability that $v_0$ is above $V_{th}$, multiplied by the probability that $v_0$ was arrived at from voltages $v_{-1}$ that were below threshold, summed over all possible subthreshold values of $v_{-1}$:
\begin{equation}
p(v_0|{\rm spike})=Z_0^{-1}H(v_0-V_{th})\displaystyle \int_{-\infty}^{V_{th}} dv_{-1}p(v_{0}|v_{-1})p(v_{-1}),
\label{eq:STdist}
\end{equation}
where $p(v_{-1})$ is the unconditioned distribution of voltages prior to the spike and is given by the steady state distribution in equation \ref{eq:LinStead}, $p(v_0|v_{-1})$ is the forward transition distribution, $H(v_0-V_{th})$ is the Heaviside function representing the probability for $v_0$ to be above threshold, and $Z_0$ is the normalization constant.

The mean voltage at the time immediately preceding the spike, $\bar{v}_{-1}$, is determined by averaging over all voltages below threshold that can transition to voltages above threshold in the next time step, and can be found from
\begin{eqnarray}
\bar{v}_{-1}&=&\displaystyle \int_{-\infty}^{\infty}\ dv_{-1}\,v_{-1}\,p(v_{-1}|{\rm spike}), \\
p(v_{-1}|{\rm spike})&=&Z^{-1} H(V_{th}-v_{-1})\displaystyle\int_{V_{th}}^{\infty}dv_0\,p(v_{-1}|v_0)p(v_0|{\rm spike}),
\end{eqnarray}
where $Z$ is the normalization constant for this distribution.  Similarly, the remaining $\bar{v}_n$ for $n\le-2$ are given by:
\begin{equation}
\bar{v}_n=\displaystyle \int_{-\infty}^{\infty} dv_n\ldots dv_{-1}\,v_{n}\,p(v_n|v_{n+1})\ldots p(v_{-1}|{\rm spike}).
\label{eq:STVdiscreet}
\end{equation}

Equation \ref{eq:STVdiscreet} is exact for a linear model but is impractical to use.  Without noticeable loss of accuracy for the simulations considered in this paper, numerous simplifications can be made.  For a discussion of approximations to $\bar{v}_0$ and $\bar{v}_{-1}$, see the appendix.

Via the central limit theorem, the $\bar{v}_n$ for $n\le -2$ can be simplified as:
\begin{equation}
\bar{v}_n\approx  \displaystyle \int_{-\infty}^{\infty} dv_n\,v_{n}\,p(v_n|\bar{v}_{n+1}).
\end{equation}
Since these are all gaussian integrals over an infinite domain, the mean value is the most likely value and so the remaining averages are arrived at recursively to give:
\begin{equation}
\bar{v}_n=\left(1-\frac{hk_{\sigma}}{\tau}\right)\bar{v}_{n+1}+\frac{hc_{\sigma}}{\tau} \quad \text{for } n\le-2.
\end{equation}
Using the fact that $\frac{h}{\tau_m}\ll 1$, this recursion relation can be solved in terms of the exponential function and gives:
\begin{equation}
\bar{v}_n=\frac{c_{\sigma}}{k_{\sigma}}+\left(\bar{v}_{-1}-\frac{c_{\sigma}}{k_{\sigma}}\right)\exp\left[\frac{k_{\sigma}(n+1)h}{\tau_m}\right] \quad \text{for}\quad n\le -1.
\end{equation}

The STA immediately follows from equation \ref{eq:STAdeflin}, and is:
\begin{equation}
\bar{s}_n=\begin{cases} 2 k_{\sigma}\left(\bar{v}_{-1}-\frac{c_{\sigma}}{k_{\sigma}}\right)\exp\left[\frac{k_{\sigma}(n+1)h}{\tau_m}\right]&n\le-1, \\ k_{\sigma}\bar{v}_0+\left(\bar{v}_0-\bar{v}_{-1}\right)\frac{\tau_m}{h}&n=0. \end{cases}
\label{eq:STAlin}
\end{equation}
For the linear model, the STA is simply given by the exponential filter up to a singular piece at the spike time that arises from requiring that threshold be crossed from below.

\subsection{Comparison to numerics}

\begin{figure}[!ht]
\centering
\includegraphics[width=4.25in]{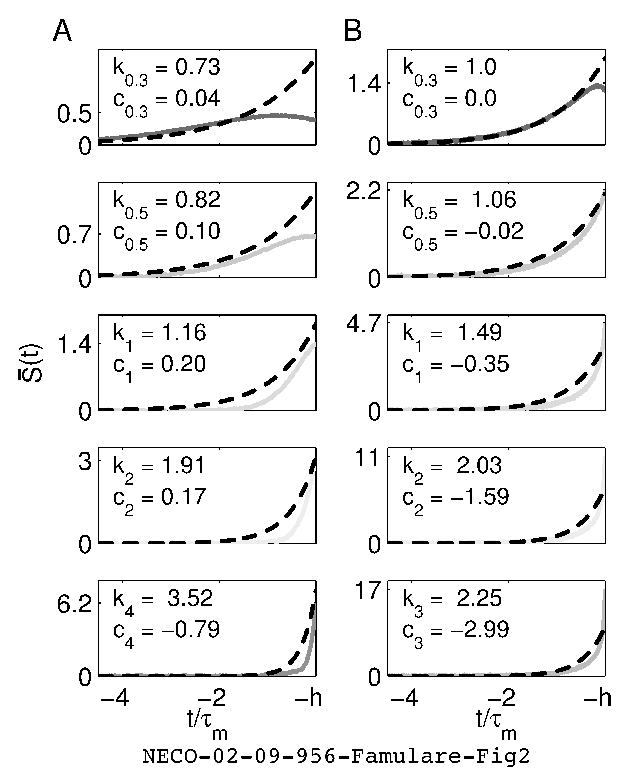}
\caption{Comparison between the numerical STA (solid) and the STA predicted by equation \ref{eq:STAlin} (dashed). (A) For the QIF, numerical results (solid) are compared to the prediction from stochastic linearization via the Energy criterion (dashed).  In this case, the $D_{KL}$ criterion predicts STA amplitudes that are too large (not shown) (B) For the EIF, numerical results (solid) are compared to the prediction from SL via the $D_{KL}$ criterion.  In this case, the Energy criterion predicts $k_{\sigma}\approx1$ for all $\sigma$ (not shown). }
\label{fig:STAhigh}
\end{figure}

\begin{figure}[!ht]
\centering
\includegraphics[width=4.25in]{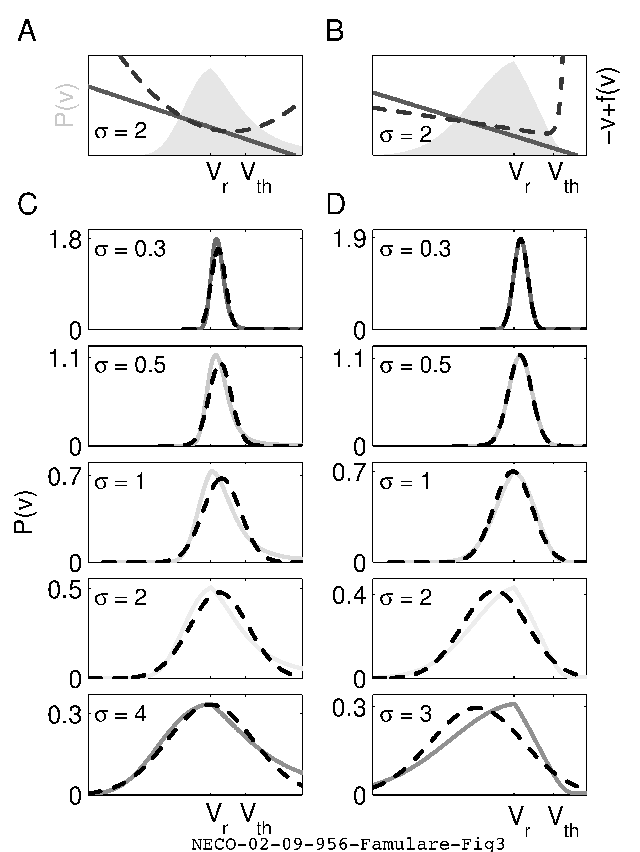}
\caption{The left column refers to the QIF and its optimization via the energy criterion, and the right, the EIF via the $D_{KL}$.  (A) This figure shows how the linear model corresponding to the energy criterion for $\sigma=2$ for the QIF corresponds to the full nonlinear model.  In gray, we see the steady-state voltage distribution.  The quadratic nonlinearity is shown dashed and the optimal linear model as determined by the energy criterion is the solid line.  We see that the linear model in this case is closely related to the average slope of the quadratic nonlinearity below threshold.  (B) In contrast, for the EIF, using the $D_{KL}$ criterion, the optimal linear model does not correspond closely with the exponential nonlinearity.  This is indicative of the fact that the adaptation of the STA in the EIF is due primarily to the spiking reset, as evinced by the STA results (see figure \ref{fig:STAhigh}).   (C,D) The steady state distributions of the (QIF,EIF) models (solid lines) are compared to their linear model approximations (dashed).  }
\label{fig:Distributions}
\end{figure}

For the QIF, the energy criterion qualitatively captures the time constant at all $\sigma$ and correctly matches the amplitude of the STA at high $\sigma$ (see figures \ref{fig:STAhigh}A and \ref{fig:Distributions}C).
Conversely, the $D_{KL}$ criterion is much less accurate.  While it too predicts qualitatively correct time constants, the amplitude of the predicted STAs is much too large (not shown).  This is because the $D_{KL}$ criterion strongly weights the effects of the after-spike reset and total probability mass, and thus biases the resting potential of the linear model too far below threshold.  Thus, the STA for the QIF is primarily determined by the form of the subthreshold nonlinearity and is less sensitive to the escape from the subthreshold domain due to spiking.  Figure \ref{fig:Distributions}A shows how the optimal linear model relates to the subthreshold nonlinearity in this case.

For the EIF, however, the energy criterion gives $k_{\sigma}\approx1$ for all $\sigma$.  While this is not surprising given the effectively linear subthreshold dynamics, it does not agree with the numerics.
The $D_{KL}$ criterion, on the other hand, applied to the EIF leads to qualitative agreement between the linear models and numerics (see figures \ref{fig:STAhigh}B and \ref{fig:Distributions}D).  This confirms the idea that the changes in the STA in the EIF can only be due to the reduction in the time spent below threshold because of spiking, and that the small amount of nonlinearity below threshold for the parameters used is not relevant except at small $\sigma$ (see figure \ref{fig:Distributions}B for further discussion). These changes in the STA are analogous to those studied previously by Paninski \citep*{Paninski2005}. 

The singular upward step at the spike time arises from the condition that threshold must be crossed from below---that $\dot{v}$ must be positive---to elicit a spike \citep*{Aguera2003a,Hong2007}.  This ``delta-function'' component, shown in figures \ref{fig:STAnum}C and \ref{fig:STAnum}D, appears here so prominently because we have chosen the spike-defining threshold to be at a voltage for which the stimulus is still relevant to spiking. This step does not vanish in the continuous-time limit.  The value of the step can be calculated approximately (see appendix) with good agreement with simulation data (see figure \ref{fig:STAdelta}).  This mode occasionally appears in experimental STAs when the spike waveform is slow (R. Mease, personal communication).  It is usually not seen because the threshold is generally drawn well into the intrinsically excitable domain of the voltage and so a condition on $\dot{v}$ does not significantly constrain the stimulus in that situation.

\section{Discussion}
\label{sec:discussion}
Due to nonlinearity, LN characterizations of neural systems show dependence on stimulus statistics, even without changes in the underlying dynamical parameters \citep*{Theunissen2000,Yu2003,Borst2005,Gaudry2007a,Gill2008,WestwickKearney}. In particular, by changing only the input standard deviation, the effective computation changes its functional form and timescale. We have explored the consequences of this for two reduced naturally-spiking neuron models, the quadratic and exponential integrate-and-fire models.  In determining the linear filter or filters characterizing the model, our work differs significantly from previous approaches \citep*{GerstnerKistler,Hong2007, Wilson2008, Badel2008a} in that the point of linearization is not taken, as is classically done, to be the equilibrium point; rather, we allow the subthreshold voltage distribution to determine the optimal point of linearization. This distribution carries information about the form of the nonlinearities, the mean firing rate, and the stimulus itself.  These properties account for changes in the effective linear model with stimulus variance. We find that despite these models' superficial similarities, different mechanisms are primarily responsible for this form of adaptation. The key difference between the models is that the QIF is nonlinear below threshold, qualitatively corresponding to a neuron with hyperpolarizing currents that are activated below threshold, while the EIF is mostly linear below threshold.  Both models have been successfully fit to neuronal data from a variety of neuron types \citep*{Izhikevich2004,Rauch2003,Badel2008c}.

Thus, both the neuron's intrinsic properties and the statistics of the background or of the driving stimulus ensemble determine the effective filtering properties of the system. This shows one means by which modulating the statistics of the input can effectively gate the transmission of different types of input or stimulus features through the system \citep*{Hasenstaub2007,Destexhe1999,Fellous2003}.  While this analysis focused on very simple model neurons, the methods we describe generalize to more complex, higher dimensional neuronal models, although analytical solutions are unlikely. These simple examples give a clear insight into intrinsic modulation of feature selectivity.

Our previous treatments of this problem \citep*{Aguera2003a,Aguera2003b,Hong2007} concentrated on the case where spikes are well-separated, so that the effects of spike history are explicitly separated from the role of the stimulus in determining the probability of generating a spike. Another approach to this problem is to include an explicit spike-history term in the generative model \citep*{GerstnerKistler,Paninski2004,Powers2005}. Here, the spike history is incorporated into the computation of features due to the effects of the mean firing rate on the steady-state distribution of threshold escape and reset. These results underscore the difficulty in inferring information about underlying biophysical parameters from the output of reverse correlation, independent of a consideration of the stimulus properties.

\section{Acknowledgments}
We thank Brian Lundstrom, Sungho Hong, Liam Paninski, and our reviewers for helpful comments and discussions.  This research is funded by the McKnight Endowment Fund for Neuroscience.

\appendix

\section*{Appendix: Approximating the singular piece of the STA}
\label{sec:deltapiece}
In numerical investigation, we find that there is a simple approximation for the average voltage at the spike time, $\bar{v}_0$, for the range of $\sigma$ considered in this paper and the use of the dynamical threshold for spike triggering. In our hands, this relationship does not seem to follow from an obvious perturbative calculation. For the QIF and EIF, we find to first order in $\sigma\sqrt{\frac{h}{\tau_m}}$:
\begin{equation}
\bar{v}_0\approx V_{th}-\frac{c_{\sigma}}{k_{\sigma}}+f\sigma\sqrt{\frac{h}{\tau_m}}
\label{eq:STV0app}
\end{equation}
where $f=0.85$ is the result of a fit to the exact integral in equation \ref{eq:STV0} for different $\sigma$, $h$, and $\tau_m$.

The exact integral for the average voltage at the time immediately preceding the spike, $\bar{v}_{-1}$, can also be approximated with an error of a few parts in a thousand.  The distribution, $p(v_{-1}|{\rm spike})$, can be approximated as:
\begin{equation}
p(v_{-1}|{\rm spike})\approx \frac{H(V_{th}-v_{-1})p(v_{-1}|\bar{v}_0)}{\displaystyle \int_{-\infty}^{V_{th}}dv_{-1}p(v_{-1}|\bar{v}_0)}
\end{equation}
To first order in $\sigma\sqrt{\frac{h}{\tau_m}}$, where $\bar{v}_0$ is given by equation \ref{eq:STV0app}, it is possible to show that:
\begin{equation}
\bar{v}_{-1}\approx V_{th}-\frac{c_{\sigma}}{k_{\sigma}}-\sigma \sqrt{\frac{h}{\tau_m}}\left(\sqrt{\frac{2}{\pi}}+f\left(\frac{2}{\pi}-1\right)\right)
\end{equation}

\begin{figure}[!ht]
\centering
\includegraphics[width=4.25in]{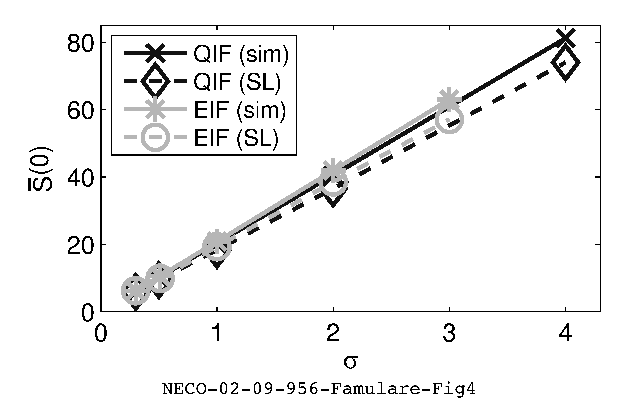}
\caption{The value of the STA in the singular component at the time of the spike, $\bar{s}(0)$, for all cases studied in this paper.  As explained in the appendix, the value is approximately linear in $\sigma$ and model-independent.    }
\label{fig:STAdelta}
\end{figure}

Taken together, this shows that the singularity in the STA at the spike time, which is given by $\dot{v}$, goes as $\sigma\sqrt{\frac{\tau_m}{h}}$.  See figure \ref{fig:STAdelta} for numerical results.

\end{document}